\documentclass[12pt]{article}
\usepackage{amssymb}

\usepackage{amsmath,amssymb,graphicx}
\usepackage{axodraw}
\usepackage{epsfig}
\usepackage{psfig}
\usepackage{citesort}

\begin{document}

\date{\today}

\renewcommand{\baselinestretch}{1.2} \vspace{0.3cm} \pagestyle{empty} 
\begin{titlepage}

\begin{flushright}
IPPP/03/19\\
DCPT/03/38\\
\today
\end{flushright}
\vspace{.3cm}

\begin{center}

\bigskip
{\Large\bf Neutrino CP phases and  lepton electric dipole moments  in 
supersymmetric theories}\\
\bigskip
\bigskip

{\large G.C. Branco$^a$\footnote{E-mail: gbranco@cfif.ist.utl.pt}, D.
Del\'{e}pine$^a$\footnote{E-mail: David.Delepine@cfif.ist.utl.pt} and S. Khalil $^{b,c}$
\footnote{E-mail: shaaban.khalil@durham.ac.uk} }

\bigskip
\emph{$^a$ Centro de F\'{\i}sica das Interac\c{c}\~{o}es Fundamentais,
Departamento de F\'{\i}sica,\\ Instituto Superior T\'{e}cnico, P-1049-001
Lisboa, Portugal\\
$^b$ IPPP, University of Durham, South Rd., Durham
DH1 3LE, U.K.\\
$^c$ Ain Shams University, Faculty of Science, Cairo, 11566, Egypt.}

\bigskip

\begin{abstract}
We analyse the dependence of the electron electric dipole moment (EDM) on 
neutrino CP violating phases in the context of supersymmetric models. 
We start by studying the supersymmetric contributions to the lepton EDM and lepton 
flavour violation processes $\tau \to \mu \gamma$ and $\mu \to e \gamma$, in 
the framework of the mass insertion approximation, showing that, due to the 
large neutrino mixing,  $\mu \to e \gamma$ 
leads to severe constraints on the relevant mass insertions. We derive model 
independent bounds on these mass insertions and show that once these bounds are 
satisfied, the present experimental limits on electron EDM do not constraint the 
neutrino phases. 

\vspace{5mm}

\end{abstract}
\end{center}
\end{titlepage}
\setcounter{page}{1} \pagestyle{plain} 

\section{Introduction}

The Standard Model (SM) of electroweak and strong interactions has had an impressive success 
when confronted with experiment. So far, we have only two pieces of evidence favouring the presence of
physics beyond the SM, namely, the experimental evidence for neutrino oscillations 
\cite{SuperK,CHOOZ,SNO,MACRO,KAMLAND} pointing towards non-vanishing neutrino masses and mixings and
the observed size of the  baryon asymmetry of the Universe (BAU). It has been 
established that the strength of the CP violation in the Standard Model is not sufficient to generate 
the cosmological baryon asymmetry of our universe, thus requiring the presence of new sources of 
$CP$ violation \cite{SM}. The most attractive mechanisms to generate the observed BAU are 
leptogenesis and electroweak baryogenesis in supersymmetric extensions of the SM. It was shown that the
supersymmetric extensions of the SM have all the necessary requirements to
generate enough BAU. In particular, the SUSY models have new sources for $CP$
violation and with a light stop, the phase transition becomes much stronger 
\cite{lightstop}. However, the bound on  the neutron EDM imposes severe constraints
on the flavour diagonal phases and they could rule out the electroweak
baryogenesis scenarios \cite{edm1}. A possible way to overcome this problem and
to generate enough BAU, without overproducing the EDMs, is to assume that SUSY $CP$ violation has 
a flavour character as in the SM \cite{ana}. 
In the leptogenesis scenario, baryon asymmetry is generated by a lepton
asymmetry arising from the out of equilibrium decay of heavy Majorana
neutrinos \cite{leptogenesis}, which is then converted into a baryon asymmetry through sphaleron 
interactions. The leptogenesis mechanism is specially attractive due to its simplicity 
and to the recent experimental results confirming neutrino oscillations and 
hence non-vanishing neutrino masses. The relation between leptogenesis and $CP$ violation
observable at low energy (in lepton EDMs or in neutrino oscillation
asymmetries) remains an open and interesting question \cite{cplowenergy-lepto}.

Until now, no $CP$ violation effects have been observed in the leptonic sector.
However, a new method for measuring lepton EDMs \cite{lamoreaux} and the prospects of 
$\nu$-factories provide the hope of having a drastic improvement in our knowledge of $CP$ 
violation in the leptonic sector, within  a few years.

In this paper, we study the impact of $CP$ violating phases in the neutrino
sector, on $CP$ violating low energy observables, in general SUSY models. In
particular, we focus on their effect on the EDM of electron and muon. In order to perform 
a model independent analysis, we use
the mass insertion approximation method which allows to parametrize the main
source of the $CP$ and flavour violation in SUSY model. In this framework, the
neutralino and chargino exchanges give the dominant contributions to the
lepton flavour violation (LFV) processes and to the EDMs.
We  derive model independent bounds on these mass
insertions and we  discuss possible constraints on the neutrino $CP$
phases from these bounds, in the  case of non-universal SUSY soft-breaking terms.
The bounds on the leptonic mass insertions provide useful tests on SUSY models and are 
complementary to those obtained in the quark sector \cite{Gabbiani96,Khalil:2001wr}. 
To illustrate these constraints,  we shall focus our interest on models where all the $CP$ 
violation is induced by the $CP$ violating phases of the leptonic mixing matrix ($U_{MNS}$), 
appearing in $W$-mediated charged current interactions.

In Ref.\cite{ellis}, the lepton EDMs have been studied in a minimal supersymmetric seesaw model with 
universality of  soft SUSY breaking terms. It was shown that,  through the renormalization group equations 
(RGE), $CP$ violating phases are  induced in the off-diagonal elements of slepton mass
matrix $m_{\tilde{L}}^{2}$ and the trilinear coupling $A_{e}$. Some of these $CP$ phases are related to the 
$CP$ phases of the lepton mixing matrix through the RGE.  Also, it was emphasized that in the case of 
non-degenerate heavy Majorana neutrinos the EDM of the  muon and the electron could be enhanced.

The paper is organized as follows. In section II, we shall introduce our notation and convention. 
The dependence of the soft SUSY breaking terms  in the $CP$ violating phases of the leptonic mixing 
matrix is extensively discussed. In section III, the analytical expressions for the lepton EDMs and 
for lepton flavour violating processes are given in terms of the mass insertions approach. In section IV, 
the bounds on the mass insertions coming from $Br(\mu \rightarrow e \gamma)$ and $d_e$ are given and  
their dependence on SUSY parameters is discussed. The recent determination of the elements of the lepton 
mixing matrix is also used to get a strong limit on the chargino contribution to  lepton flavour violating 
processes such as $Br(\tau \rightarrow \mu,e \gamma)$. In the section V, we shall discuss the dependence 
of the lepton EDM on the $CP$  violating phases appearing in the leptonic mixing matrix. Particular emphasis 
is given to the Majorana phases dependence. We show that if there is no SUSY $CP$ violating phases, 
whatever is the structure of the soft SUSY breaking terms,  the lepton EDMs only depend on the Dirac 
$CP$ violating phase of $U_{MNS}$. As an illustrative example, we studied the case of Hermitian 
Yukawa couplings. In section VI, we summarize 
our main results and present our conclusions.


\section{Supersymmetric model with right--handed neutrinos}

The seesaw mechanism\cite{seesaw} provides a natural explanation for the smallness of neutrino masses 
which are of order $v^2/M_R$ where $v$ stands for the scale of electroweak symmetry breaking and $M_R$ 
denotes the right-handed neutrino mass. Since the right-handed Majorana neutrino mass term is $SU(2)_L 
\otimes U(1)_Y \otimes SU(3)_{QCD}$ invariant, $M_R$ can have a value much larger than $v$. 
Supersymmetry can play an important r\^ole in  ensuring the stability of the  hierarchy between
the weak scale and right--handed neutrino scale. We consider the
supersymmetric standard model with right-handed neutrinos, which is
described by the superpotential 
\begin{equation}
W=-\mu H_{1}H_{2}+Y_{eij}E_{i}^{c}L_{j}H_{1}+Y_{\nu ij}N_{i}^{c}L_{j}H_{2}+{%
\frac{1}{2}}Y_{rij}N_{i}^{c}N_{j}^{c}R
\end{equation}
where $i,j=1\ldots 3$ are generation indices and the superfields $E^{c}$, $%
L=(N,E)$, $N^{c}$ contain the leptons $e_{R}^{c}$, $(\nu _{L},e_{L})$, $\nu
_{R}^{c}$, respectively. The expectation values of the Higgs multiplets $%
H_{1}$ and $H_{2}$ generate ordinary Dirac mass terms for quarks and leptons, and
the expectation value of the singlet Higgs field $R$ yields the Majorana
mass matrix of the right-handed neutrinos, $M_{Rij}=Y_{rij}\langle R\rangle $%
. In general, the Majorana neutrino mass matrix is a complex symmetric
matrix. At low energy and after the decoupling of the heavy neutrinos, the
effective superpotential is given by 
\begin{equation}
W_{\mathrm{eff}}=(Y_{\nu ij})_{\mathrm{eff}%
}L_{i}L_{j}H_{2}^{2}+Y_{eij}E_{i}^{c}L_{j}H_{1},
\end{equation}
where  $(Y_{\nu })_{\mathrm{eff}%
}=-Y_{\nu }^{T}M_{R}^{-1}Y_{\nu }$ and the light neutrino masses are given
by $M_{\nu }=(Y_{\nu })_{\mathrm{eff}}\langle H_{2}^{0}\rangle ^{2}$. Since  $M_{\nu }$ is a 
symmetric matrix, it can be diagonalized by
a unitary transformation. In addition, the relevant soft SUSY breaking terms are in general  given by 
\begin{eqnarray}
L_{soft} &=&-{\widetilde{m}}_{lij}^{2}L_{i}^{\dagger }L_{j}-{\widetilde{m}}%
_{eij}^{2}E_{i}^{c\dagger }E_{j}^{c}  \label{soft} \\
&&+Y_{eij}^{A}E_{i}^{c}L_{j}H_{1}+Y_{\nu
ij}^{A}N_{i}^{c}L_{j}H_{2}+c.c.+\ldots \;,  \nonumber
\end{eqnarray}
where $L=(N_{L},E_{L})$ and $E^{c}\equiv E_{R}^{*}$ refer to the scalar
partners of $(\nu _{L},e_{L})$ and $e_{R}^{c}$ respectively and $Y_{e,\nu
ij}^{A}\equiv $ $A_{e,\nu ij}Y_{e,\nu ij}$. Using the seesaw mechanism to
explain the smallness of neutrino masses, we assume that the right-handed
neutrino masses $M_{R}$ are much larger than the Fermi scale $v$. One then
easily verifies that all mixing effects on light scalar masses caused by the
right-handed neutrinos and their scalar partners are suppressed by $\mathcal{%
O}(v/M_{R})$, and therefore negligible. 

Now we present the general expressions for the slepton mass mass matrices in
the super-MNS basis which, in analogue to the super--CKM basis in the quark
sector, is defined as follows. Given the Yukawa matrices, we perform unitary
transformations of the lepton superfields $N_{L}$ and $E_{L,R}$ such that
the lepton mass matrices take diagonal forms: 
\begin{eqnarray}
N_{L} &\to &V_{L}^{\nu }N_{L}\;,  \nonumber \\
E_{L,R} &\to &V_{L,R}^{e}E_{L,R}\;,
\end{eqnarray}
with $Y_{eff}^{\nu }\rightarrow (V_{L}^{\nu })^{T}Y_{eff}^{\nu }V_{L}^{\nu }=%
\mathrm{diag}(h_{\nu _{e}},h_{\nu _{\mu }},h_{\nu _{\tau }})$ and $%
Y^{e}\rightarrow (V_{R}^{e})^{\dagger }Y^{e}V_{L}^{e}=\mathrm{diag}%
(h_{e},h_{\mu },h_{\tau })$. In this basis, the leptonic charged current
interactions is given by 
\begin{equation}
-\frac{g}{\sqrt{2}}\left( \overline{l}_{Li}\gamma ^{\mu }(V_{L}^{e\dagger
}V_{L}^{\nu })_{ij}\nu _{j}W_{\mu }+h.c.\right) 
\end{equation}
with $g$, the weak $SU(2)_{L}$ gauge coupling. The lepton flavour mixing
matrix is then given by 
\begin{equation}
U_{MNS}=V_{L}^{e\dagger }V_{L}^{\nu }.
\end{equation}
Both  $V_{L}^{e}$ and $V_{L}^{\nu }$ are unitary  matrices which can be parametrised as
\begin{equation}
V_{L}^{e,\nu }\equiv P_{e,\nu }V_{\delta }^{e,\nu }Q_{e,\nu }
\end{equation}
where $P_{e,\nu }$ $\equiv \mathrm{diag}(e^{i\alpha _{1}^{e,\nu
}},e^{i\alpha _{2}^{e,\nu }},e^{i\alpha _{3}^{e,\nu }})$,  $Q_{e,\nu }$ $%
\equiv \mathrm{diag}(1,e^{i\beta _{1}^{e,\nu }},e^{i\beta _{2}^{e,\nu }})$
and $V_{\delta }^{e,\nu }$ are unitary matrices which only contain one $CP$
violating phase. So, $U_{MNS}$ can be rewritten as 
\begin{eqnarray}
U_{MNS} &=&Q_{e}^{\dagger }\underbrace{V_{\delta }^{e\dagger }P_{e}^{\dagger
}P_{\nu }V_{\delta }^{\nu }}Q_{\nu } \\
&=&Q_{e}^{\dagger }\left( P^{\dagger }U_{\delta }Q\right) Q_{\nu } \\
&\equiv &P_{L}^{\dagger }U_{\delta }P_{M}  \label{U_MNS_phases}
\end{eqnarray}
where $U_{\delta }$ is a unitary transformation parametrised by three angles
and one $CP$ violating phase, similar to $V_{CKM}$ quark mixing matrix. $P$
and $Q$ are diagonal phase matrices similar to $P_{e,\nu }$ and 
$Q_{e,\nu }$, respectively. As one can see from eq.(\ref{U_MNS_phases}), in general $%
U_{MNS}$  contains six $CP$ violating phases but as
it is the case in the SM, it is always possible to redefine the $E_{L}$
superfields by a diagonal unitary transformation such that three phases
contained in $Q_{e}^{\dagger }P^{\dagger }\equiv P_{L}^{\dagger }$ are removed and an
equivalent transformation on $E_{R}$ superfields in order to keep the
charged lepton masses real. So, only three phases of $U_{MNS}$ are physical:
one coming from the unitary matrix $U_{\delta }$ which contains one $CP$
violating phase (this phase is usually called the Dirac phase) , and a
diagonal unitary matrix $P_{M}\equiv (1,e^{i\alpha },e^{i\beta })\equiv
QQ_{\nu }$ which contains the two Majorana phases. In this basis, $U_{MNS}$
is given by 
\begin{equation}
U_{MNS}=U_{\delta }P_{M}  \label{U_MNS}
\end{equation}
with $P_{M}\equiv \mathrm{diag}(e^{i\phi _{j}^{M}})_{j=1..3}$ with $\phi
_{1}^{M}=0$ and $\phi _{2,3}^{M}$ are the usual Majorana phases.

In this super--MNS basis, the low-energy sneutrino and charged slepton mass
matrices are given by 
\[
M_{\tilde{e}}^{2}=\left( 
\begin{array}{cc}
\left( M_{\tilde{e}}^{2}\right) _{LL} & \left( M_{\tilde{e}}^{2}\right) _{LR}
\\ 
\left( M_{\tilde{e}}^{2}\right) _{RL} & \left( M_{\tilde{e}}^{2}\right) _{RR}
\end{array}
\right) ,
\]
where 
\begin{eqnarray}
\left( M_{\tilde{e}}^{2}\right) _{LL} &=&P_{L}V_{L}^{e\dagger }\tilde{m}%
_{l}^{2}V_{L}^{e}P_{L}^{\dagger }+m_{l}^{2}+\frac{m_{Z}^{2}}{2}(1-2\sin
^{2}\theta _{W})\cos {2\beta },  \nonumber \\
\left( M_{\tilde{e}}^{2}\right) _{RR} &=&P_{L}V_{R}^{e\dagger }\tilde{m}%
_{e}^{2}V_{R}^{e}P_{L}^{\dagger }+m_{l}^{2}-\frac{m_{Z}^{2}}{\sin ^{2}{%
\theta _{W}}}\cos {2\beta },  \nonumber \\
\left( M_{\tilde{e}}^{2}\right) _{RL} &=&\left( M_{\tilde{e}}^{2}\right)
_{LR}^{\dag }=-\mu ~m_{l}~\tan \beta +v\cos \beta ~P_{L}(V_{R}^{e})^{\dagger
}Y_{e}^{A}V_{L}^{e}P_{L}^{\dagger }~,  \label{sqmass1}
\end{eqnarray}
and 
\begin{equation}
\left( M_{\tilde{\nu}}^{2}\right) _{LL}=V_{L}^{\nu \dagger }{\tilde{m}%
_{l}^{2}}V_{L}^{\nu }+\frac{m_{Z}^{2}}{2}\cos {2\beta +}v\sin \beta
V_{L}^{\nu \dagger }\left( Y^{\nu \dagger }Y^{\nu }\right) V_{L}^{\nu }.
\label{sqmass2}
\end{equation}
where $m_{l}$ is the diagonal charged lepton mass matrices, $m_{Z}$ is the
mass of the $Z$ gauge boson, ${\theta _{W}}$ is the usual Weinberg angle of
the weak interactions and $\tan \beta \equiv \left\langle
H_{2}\right\rangle /\left\langle H_{1}\right\rangle $.
In the last equation, we have kept the contribution to $M_{\tilde{\nu}}^{2}$
proportional to the Dirac mass of the neutrinos because in general the
unitary transformation which diagonalised $Y_{eff}^{\nu }$ doesn't
necessarily diagonalise $Y^{\nu \dagger }Y^{\nu }$. Recall that, due to very
heavy Majorana masses, the $RR$ and $LR$ contributions to the sneutrino mass
matrix are suppressed. As it can be seen from eqs.(\ref{sqmass1}-\ref
{sqmass2}), for non-universal soft breaking terms, the dependence on the $CP$
violating phases coming from $Y_{eff}^{\nu }$ and $Y^{e}$ is very involved
and in general, their contributions can not be distinguished from the $CP$
violating phases arising from the soft breaking terms. Of course, we
could have started to work from the beginning in the mass eigenstates basis
for the leptons, where the $U_{MNS}$ matrix contains only three phases.
But usually, in models with non-universal SUSY soft breaking
terms and flavour symmetry, the textures of the soft-breaking terms ( $%
\tilde{m}_{l}^{2},\tilde{m}_{e}^{2}$ and $Y_{e}^{A}$) are given in a weak
basis where in general $Y^{e}$ and $Y_{eff}^{\nu }$ are not diagonal (see
for instance, refs.\cite{vissani98}). 

In this paper, we shall be interested in studying the effects of $CP$-violating phases arising 
from the lepton mass matrices ($ \approx Y^e$ and $Y^{\nu}_{eff}$) on charged leptons EDM's 
taking into account the bounds on flavour changing lepton decays.

Since charged leptons are in and out states in $l_i^- \rightarrow l_j^+ \gamma$ or in lepton EDM's, 
the computation of the  contribution to these processes involving the supermultiplets $E_{L,R}$ 
has to be performed in the charged lepton mass eigenstates basis. But for supermultiplets $N_L$, 
as they only contribute to charged lepton decays and EDM's through loop contribution, the 
computation of their contribution to these processes can be done in any weak basis for the 
supermultiplets $N_L$. As a corollary, it means that these processes are independent of 
$V_L^{\nu}$. Thus, we can directly conclude that if the textures for $Y^e$ and 
$Y^{\nu}_{eff}$ are such that $U_{MNS}$ is dominated by $V_{L}^{\nu }$, there is no
effect of the neutrino $CP$ phases on these different processes for any
kind of textures for the SUSY soft breaking terms. But the situation is completely different 
if the textures for the Yukawa couplings give $V_{L}^{e\dagger }\simeq $ $V_{L}^{\nu }$ or if $%
V_{L}^{\nu }\simeq \mathbf{1}$. Indeed, in such a case, as one can see from
eqs(\ref{sqmass1}-\ref{sqmass2}), the neutrino $CP$ phases dependence of the
SUSY soft breaking terms can be explicitly studied. In the following part of
the paper, we shall assume that $V_{L}^{\nu }\simeq \mathbf{1}$. This means
that we shall assume that the origin of the large lepton mixing angles comes
from the charged lepton Yukawa couplings. In the ``super-MNS'' basis where
all lepton masses are real and $U_{MNS}$ only contains three phases, one has
for $V_{L}^{\nu }\simeq \mathbf{1,}$%
\begin{equation}
U_{MNS}\simeq P_{L}V_{L}^{e\dagger }  \label{definition_U_MNS}
\end{equation}
with $P_{L}\simeq Q_{e}$. 

Moreover in the ``super-MNS'' basis, the couplings of lepton and slepton
states to the neutralinos are flavour diagonal and all the source of flavour
mixing are inside the off--diagonal terms of the slepton mass matrix. These
terms are denoted by $(\Delta _{AB}^{l})^{ij}$, where $A,B=(L,R)$ for $%
l\equiv e$ and $A,B=L$ for $l\equiv \nu $ and $i,j=1,2,3$ denote the flavour
indices. The slepton propagator is expanded as a series of the dimensionless
quantity $(\delta _{AB}^{l})_{ij}=(\Delta _{AB}^{l})^{ij}/\tilde{m}^{2}$,
where $\tilde{m}^{2}$ is an average slepton mass. 
\begin{equation}
\langle \tilde{l}_{A}^{i}\tilde{l}_{B}^{j}\rangle =i\left( k^{2}\mathbf{1}-%
\tilde{m}^{2}\mathbf{1}-\Delta _{AB}^{l}\right) _{ij}^{-1}\simeq \frac{%
i\delta _{ij}}{k^{2}-\tilde{m}^{2}}+\frac{i(\Delta _{AB}^{l})_{ij}}{(k^{2}-%
\tilde{m}^{2})^{2}}+\mathcal{O}(\Delta ^{2}),
\end{equation}
where $l=\nu ,e$ denote the neutrino and charged lepton sectors
respectively. As mentioned, $A,B$ stand for $L$ with neutrino sector and $L,R
$ for the charged sector, $i,j=1,2,3$ are the flavour indices, $\mathbf{1}$
is the unit matrix, and $\tilde{m}$ is the average slepton mass. This
method, known as mass insertion approximation, allows to parametrize, in
a model independent way, the flavour violation in supersymmetric theories. 
It is also worth mentioning that since $\tilde{m}_{l}^{2}$ is Hermitian,
the mass matrices $(M_{\tilde{e},\tilde{\nu}}^{2})_{LL}$ and $(M_{\tilde{e}%
}^{2})_{RR}$ in eqs.(\ref{sqmass1}, \ref{sqmass2}) are also Hermitian in 
the ``super-MNS'' basis. Therefore, the LL mass insertions $%
(\delta _{LL}^{l,\tilde{\nu}})=1/\tilde{m}^{2}(M_{\tilde{e},\tilde{\nu}%
}^{2})_{LL}$ and $(\delta _{RR}^{l})=1/\tilde{m}^{2}(M_{\tilde{e}}^{2})_{RR}$
are also Hermitian. 

In the ``super-MNS'' basis and assuming the decoupling of the right-handed
neutrino scalars, the Lagrangian describing the interaction between the charginos and
the leptons and their partners needed to compute the chargino contributions
to the lepton EDM and LFV is given by 
\begin{equation}
\mathcal{L}_{e\tilde{\nu}\chi ^{+}}=\sum_{k}\sum_{a,b}~\Big(%
-~gV_{k1}(U_{MNS})_{ab}~\bar{e}_{L}^{a}~(\chi _{k}^{+})^{*}~\tilde{\nu}%
_{L}^{b}+U_{k2}^{*}~[Y_{e}^{\mathrm{diag}}.U_{MNS}]_{ab}~\bar{e}%
_{R}^{a}~(\chi _{k}^{+})^{*}~\tilde{\nu}_{L}^{b}~\Big),  \label{chargino}
\end{equation}
where the indices $a,b$ and $k$ label flavour and chargino mass eigenstates
respectively and $V$, $U$ are the chargino mixing matrices defined by 
\begin{equation}
U^{*}M_{\chi ^{+}}V^{-1}=\mathrm{diag}(m_{\chi _{1}^{+}},m_{\chi _{2}^{+}}),
\end{equation}
and 
\begin{equation}
M_{\chi ^{+}}=\left( 
\begin{array}{cc}
M_{2} & \sqrt{2}M_{W}\sin \beta  \\ 
\sqrt{2}M_{W}\cos \beta  & -\mu 
\end{array}
\right) \;.
\end{equation}
The relevant Lagrangian for the neutralino contributions is given by 
\begin{eqnarray}
\mathcal{L}_{e\widetilde{e}\chi ^{0}} &=&\sum_{k}\sum_{a}~\Big(g\frac{%
(N_{k2}+\tan \theta _{W}N_{k1})}{\sqrt{2}}\bar{e}_{L}^{a}~(\tilde{\chi}%
_{k}^{0})^{*}~\widetilde{e}_{L}^{a}-N_{k3}\left( Y_{e}^{\mathrm{diag}%
}\right) _{aa}~\bar{e}_{L}^{a}~(\tilde{\chi}_{k}^{0})^{*}~\widetilde{e}%
_{R}^{a}  \nonumber \\
&&-g\sqrt{2}\tan \theta _{W}N_{k1}^{*}\bar{e}_{R}^{a}~(\tilde{\chi}%
_{k}^{0})^{*}~\widetilde{e}_{R}^{a}-N_{k3}^{*}~\left( Y_{e}^{\mathrm{diag}%
}\right) _{aa}\bar{e}_{R}^{a}~(\tilde{\chi}_{k}^{0})^{*}~\widetilde{e}%
_{L}^{a}\Big).  \label{neutralino}
\end{eqnarray}
where the matrix $N$ is defined as the $4\times 4$ rotation matrix which
diagonalized the neutralino mass matrix $M_{N}$, 
\begin{equation}
N^{*}M_{N}N^{-1}=diag(m_{\chi _{1}^{0}},m_{\chi _{2}^{0}},m_{\chi
_{3}^{0}},m_{\chi _{4}^{0}}).
\end{equation}
$M_{N}$ is given by 
\begin{equation}
M_{N}=\left( 
\begin{array}{cccc}
M_{1} & 0 & -m_{Z}\sin \theta _{W}\cos \beta  & m_{Z}\sin \theta _{W}\sin
\beta  \\ 
0 & M_{2} & m_{Z}\cos \theta _{W}\cos \beta  & -m_{Z}\cos \theta _{W}\sin
\beta  \\ 
-m_{Z}\sin \theta _{W}\cos \beta  & m_{Z}\cos \theta _{W}\cos \beta  & 0 & 
\mu  \\ 
m_{Z}\sin \theta _{W}\sin \beta  & -m_{Z}\cos \theta _{W}\sin \beta  & \mu 
& 0
\end{array}
\right) ,
\end{equation}
with $M_{1,2}$ are respectively the $U(1)_{Y}$ and $SU(2)_{L}$ gaugino soft
masses. 


\section{Supersymmetric contributions to EDM and LFV}

\subsection{Electric dipole moment of charged leptons}

The effective Hamiltonian for the EDM of the charged leptons $l$ can be written as
\begin{equation}
H_{\mathrm{eff}}^{\mathrm{EDM}}=C_{1}\mathcal{O}_{1}+h.c.,
\end{equation}
where $C_{1}$ and $\mathcal{O}_{1}$ are the Wilson coefficient and the
electric dipole moment operator respectively. The operator $\mathcal{O}_{1}$
is given by 
\begin{equation}
\mathcal{O}_{1}=-\frac{i}{2}\bar{l}\sigma _{\mu \nu }\gamma _{5}lF^{\mu \nu
}.
\end{equation}
The supersymmetric contributions to the Wilson coefficient of the charged
lepton result from the one loop penguin diagrams with neutralino and
chargino exchange (figs.(1.a),(1.b)).\newline

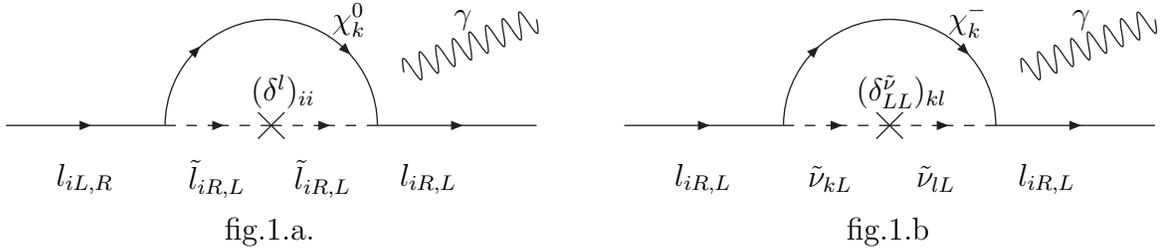
\begin{figure}[t]
\begin{center}
{\normalsize 
\begin{picture}(200,100)(65,0)
\ArrowArcn(150,50)(40,180,90)
\ArrowArcn(150,50)(40,90,0)
\ArrowLine(50,50)(110,50)
\ArrowLine(190,50)(250,50)
\DashArrowLine(110,50)(150,50){5}
\Line(145,45)(155,55)
\Line(145,55)(155,45)
\DashArrowLine(150,50)(190,50){5}
\Photon(200,70)(250,90){5}{8}
{\Text(80,30)[]{$l_{iL,R}$}}
{\Text(210,30)[]{$l_{iR,L}$}}
{\Text(130,30)[]{$\tilde{l}_{iR,L}$}}
{\Text(170,30)[]{$\tilde{l}_{iR,L}$}}
{\Text(180,90)[]{$\chi^0_k$}}
{\Text(225,90)[]{$\gamma$ }}
\Text(157,63)[]{$(\delta^l)_{ii}$ }
\Text(150,10)[]{fig.1.a.}
\end{picture}
\begin{picture}(200,100)(35,0)
\ArrowArcn(150,50)(40,180,90)
\ArrowArcn(150,50)(40,90,0)
\ArrowLine(50,50)(110,50)
\ArrowLine(190,50)(250,50)
\DashArrowLine(110,50)(150,50){5}
\Line(145,45)(155,55)
\Line(145,55)(155,45)
\DashArrowLine(150,50)(190,50){5}
\Photon(200,70)(250,90){5}{8}
{\Text(80,30)[]{$l_{iR,L}$}}
{\Text(210,30)[]{$l_{iR,L}$}}
{\Text(130,30)[]{$\tilde{\nu}_{kL}$  }}
{\Text(170,30)[]{$\tilde{\nu}_{lL}$  }} 
{\Text(180,90)[]{$\chi^{-}_k$}}
{\Text(225,90)[]{$\gamma$ }}
\Text(157,63)[]{$(\delta^{\tilde{\nu}}_{LL})_{kl}$ }
\Text(120,90)[]{ }
\Text(150,10)[]{fig.1.b}
\end{picture}
}
\end{center}
\caption{ The neutralino (fig.1.a) and chargino contributions (fig.1.b) to
the charged lepton EDM in ``Super-MNS'' basis. For neutralino (chargino)
diagram, the photon line has respectively to be attached to the scalar
(fermion) line of the loop. The cross represents the mass insertions.}
\end{figure}
The lepton EDM is given by 
\begin{equation}
d_{l}/e=\mathrm{Im}\left[ C_{l}^{\chi ^{+}}+C_{l}^{\chi ^{0}}\right] ,
\end{equation}
where $e$ is the electron electric charge. In the framework of mass
insertion approximation, we find that the neutralino contribution to the
above Wilson coefficient is given by 

\begin{eqnarray}
C_{l_{i}}^{\chi ^{0}} &=&\frac{\alpha _{W}}{4\pi }\sum_{a=1}^{4}\frac{1}{%
m_{\chi _{a}^{0}}}f_{1}(x_{a})\Big[ \mbox{ tan}\theta
_{W}N_{a1}^{*}(N_{a2}^{*}+\mbox{ tan}\theta _{W}N_{a1}^{*})~(\delta
_{RL}^{l})_{ii}  \nonumber \\
&&-N_{a3}^{*}N_{a1}^{*}\mbox{ tan}\theta _{W}\frac{m_{l_{i}}}{m_{W}%
\mbox{
cos}\beta }~(\delta _{RR}^{l})_{ii}  \nonumber \\
&&-N_{a3}^{*}(N_{a2}^{*}+\mbox{ tan}\theta _{W}N_{a1}^{*})\frac{m_{l_{i}}}{%
2m_{W}\mbox{ cos}\beta }~(\delta _{LL}^{l})_{ii}\Big]
\end{eqnarray}
where $x_{a}=\tilde{m}^{2}/m_{\tilde{\chi}_{a}^{0}}^{2}$ and the function $%
f_{1}(x)$ is given by 
\begin{equation}
f_{1}(x)=\frac{x\left( 5-4x-x^{2}+2(1+2x)\log x\right) }{2(1-x)^{4}}.
\end{equation}
It is worth mentioning that in minimal supergravity model, the lightest
neutralino leads to the dominant contribution to $C_{l_{i}}^{\chi _{j}^{0}}$%
. However, in general supersymmetric models other neutralino exchanges could
give also significant contributions to $C_{l_{i}}^{\chi _{j}^{0}}$. 

Calculating the chargino contribution to the Wilson
coefficient of the charged lepton $l_{i}$, in the mass insertion
approximation, one obtains the following expression for the Wilson coefficient $C_{l_{i}}^{\tilde{\chi}%
^{+}}$: 
\begin{eqnarray}
C_{l_{i}}^{\tilde{\chi}^{+}} &=&\frac{\alpha _{W}}{4\pi }\frac{m_{l_{i}}}{%
\sqrt{2}m_{W}\mbox{
cos}\beta }\sum_{j,k=1}^{3}\left( \left( U_{MNS}\right) _{ij}(\delta _{LL}^{%
\tilde{\nu}})_{jk}(U_{MNS}^{\dagger })_{ki}\right) \times   \nonumber \\
&&\sum_{a=1}^{2}\frac{1}{m_{\tilde{\chi}_{a}^{+}}}%
U_{a2}^{*}V_{a1}^{*}f_{2}(x_{a})  \label{dl}
\end{eqnarray}
where $m_{\tilde{\chi}_{a}}$ is the chargino mass and $x_{a}=\tilde{m}%
^{2}/m_{\tilde{\chi}_{a}^{+}}^{2}$. The loop function $f_{2}(x)$ is given by 
\begin{equation}
f_{2}(x)=\frac{x}{2(1-x)^{4}}(5x^{2}-4x-1-2x(2+x)\mbox{ ln}x)
\end{equation}
\subsection{Lepton flavour violation, $l_{i}\rightarrow l_{j}\gamma $ }

The experimental bounds on the lepton flavour violating decays of charged
leptons, in particular the $\mu \to e\gamma $, impose strong constraints on
the absolute values of the relevant mass insertions. As we will show in the
next section, these constraints have important consequences on the
prediction of the lepton EDM results. Therefore in our analysis we have to
take the effect of these decays into account. These processes
receive contributions from chargino and neutralino exchanges. Assuming $%
m_{l_{i}}\gg m_{l_{j}}$, the amplitude to $l_{i}\rightarrow l_{j}\gamma $
can be written as 
\begin{equation}
M_{l_{i}\rightarrow l_{j}\gamma }\equiv e\epsilon _{\alpha }^{*}(q)\left( 
\overline{u}_{l_{j}}i\sigma ^{\alpha \beta }q_{\beta }P_{R}u_{l_{i}}\right)
\left( A_{Rji}^{C}+A_{Rji}^{N}\right) +(L\leftrightarrow R)
\end{equation}
where $A^{C,N}$ denote the chargino and neutralino contributions, respectively. 
The neutralino contributions are given by 
\begin{eqnarray}
A_{Rji}^{N} &=&\frac{\alpha _{W}}{4\pi }m_{l_{i}}\sum_{a=1}^{4}\frac{1}{m_{%
\tilde{\chi}_{a}^{0}}^{2}}\Big[ \Big(\frac{m_{\tilde{\chi}_{a}^{0}}}{2m_{W}%
\mbox{cos}\beta }N_{a3}(N_{a2}+\tan \theta _{W}N_{a1})f_{1}(x_{a})  \nonumber
\\
&+&\frac{(N_{a2}+\tan \theta _{W}N_{a1})^{2}}{2}f_{3}(x_{a})\Big) (\delta
_{LL}^{l})_{ji}  \nonumber \\
&+&\frac{m_{\tilde{\chi}_{a}^{0}}}{m_{l_{i}}}\tan \theta
_{W}N_{a1}(N_{a2}+\tan \theta _{W}N_{a1})~f_{1}(x_{a})(\delta _{LR}^{l})_{ji}%
\Big]
\end{eqnarray}
\begin{eqnarray}
A_{Lji}^{N} &=&\frac{\alpha _{W}}{4\pi }m_{l_{i}}\sum_{a=1}^{4}\frac{1}{m_{%
\tilde{\chi}_{a}^{0}}^{2}}\Big[ \Big(-\frac{m_{\tilde{\chi}_{a}^{0}}}{m_{W}%
\mbox{cos}\beta }\tan \theta _{W}N_{a1}^{*}N_{a3}^{*}f_{1}(x_{a})  \nonumber
\\
&&+2\tan ^{2}\theta _{W}\left| N_{a1}\right| ^{2}f_{3}(x_{a})\Big) (\delta
_{RR}^{l})_{ji}  \nonumber \\
&&+\frac{m_{\tilde{\chi}_{a}^{0}}}{m_{l_{i}}}\tan \theta
_{W}N_{a1}^{*}(N_{a2}^{*}+\tan \theta _{W}N_{a1}^{*})f_{1}(x_{a})(\delta
_{RL}^{l})_{ji}\Big]
\end{eqnarray}
with $f_{3}(x)$ given by 
\begin{equation}
f_{3}(x)=\frac{x(-17+9x+9x^{2}-x^{3}-6(1+3x)\mbox{ log}x)}{12(x-1)^{5}}
\end{equation}

For the chargino contribution one obtains:
\begin{eqnarray}
A_{Rji}^{C} &=&\frac{\alpha _{W}}{4\pi }m_{l_{i}}\left( U_{MNS}\right)
_{jh}(\delta _{LL}^{\tilde{\nu}})_{hk}(U_{MNS}^{\dagger })_{ki}\times 
\label{chargino_mu_decays} \\
&&\sum_{a=1}^{2}\frac{1}{m_{\tilde{\chi}_{a}^{+}}^{2}}\left( \frac{m_{\tilde{%
\chi}_{a}^{+}}}{\sqrt{2}m_{W}\mbox{
cos}\beta }U_{a2}V_{a1}f_{4}(x_{a})-\left| V_{a1}\right|
^{2}f_{5}(x_{a})\right) .  \nonumber
\end{eqnarray}
\begin{equation}
A_{Lji}^{C}=O(m_{l_{j}})
\end{equation}
The one-loop functions are given as 
\begin{eqnarray}
f_{4}(x) &=&\frac{x}{2(1-x)^{4}}(5x^{2}-4x-1-2x(2+x)\mbox{ ln}x) \\
f_{5}(x) &=&\frac{x}{6(1-x)^{5}}(x^{3}+9x^{2}-9x-1-6x(1+x)\mbox{ ln}x).
\end{eqnarray}

The branching ratio of $\mu \rightarrow e\gamma $ can be expressed as 
\begin{equation}
Br(\mu \rightarrow e\gamma )=384\pi ^{3}\alpha \frac{v^{4}}{m_{\mu }^{2}}%
\left( \left| A_{R12}^{C}+A_{R12}^{N}\right| ^{2}+(R\leftrightarrow
L)\right) 
\end{equation}
with $v=(8G_{F}^{2})^{-1/4}\simeq 174$ GeV and $\alpha =e^{2}/4\pi $. Based
on these results on $\mu \rightarrow e\gamma $, one can immediately write
down the rate for the process $\tau \rightarrow \mu ,e\gamma $. Using $%
\Gamma _{\tau }\simeq 5(m_{\tau }/m_{\mu })^{5}\Gamma _{\mu }$, one obtains
for the branching ratios, 
\begin{eqnarray}
Br(\tau  &\rightarrow &\mu \gamma )=\frac{384}{5}\pi ^{3}\alpha \frac{v^{4}}{%
m_{\tau }^{2}}\left( \left| A_{R23}^{C}+A_{R23}^{N}\right|
^{2}+(R\leftrightarrow L)\right)  \\
Br(\tau  &\rightarrow &e\gamma )=\frac{384}{5}\pi ^{3}\alpha \frac{v^{4}}{%
m_{\tau }^{2}}\left( \left| A_{R13}^{C}+A_{R13}^{N}\right|
^{2}+(R\leftrightarrow L)\right) 
\end{eqnarray}


\section{Constraints from $BR(\mu \to e\gamma )$ and electron EDM}

In this section we present our results for the bounds on $(\delta
_{AB}^{l})_{ij}$ and $(\delta _{LL}^{\tilde{\nu}})_{ij}$ which come
respectively from the neutralino and chargino contributions to $BR(\mu \to
e\gamma )$ and electron EDM. As it is well known, until now, no lepton flavour
violating processes or electric dipole moments of lepton have been
experimentally observed. So we have only bounds on these different
processes. As can be seen from table I, where we summarize the present
experimental status, the strongest bounds are related to the electron
EDM and the $BR(\mu \to e\gamma )$. 

\begin{center}
\begin{tabular}{|c|c|}
\hline
& present bound \\ \hline
$d_{e}$ & $<4.3\times 10^{-27}e$ cm\cite{commins1994} \\ \hline
$d_{\mu }$ & $(3.7\pm 3.4)\times 10^{-18}e$ cm\cite{g-2} \\ \hline
$d_{\tau }$ & $<3.1\times 10^{-16}e$ cm\cite{acciarri1998} \\ \hline
$Br(\mu \rightarrow e\gamma )$ & $<1.2\times 10^{-11}$\cite{mega} \\ \hline
$Br(\tau \rightarrow \mu \gamma )$ & $<1.1\times 10^{-6}$\cite{CLEO} \\ 
\hline
$Br(\tau \rightarrow e\gamma )$ & $<2.7\times 10^{-6}$\cite{PDG} \\ \hline
\end{tabular}
\end{center}

Table 1: The current experimental bounds on the LVF processes and lepton
EDMs.\newline

An experiment aimed to reach the sensitivity for $Br(\mu \rightarrow e\gamma
)$ of $10^{-14}$ has been proposed at PSI\cite{PSI}, and the stopped-muon
experiment that could take place at neutrino factories could reach $Br(\mu
\rightarrow e\gamma )$ $\sim 10^{-15}$\cite{aysto}. The $B$-factories as
Belle and LHC should be able to improve $Br(\tau \rightarrow \mu \gamma )$
by typically one order of magnitude. But the most significant experimental
improvements could be expected from the electric dipole moment for the
electron and the muon. Indeed, recently, it has been proposed that one
could improve by six orders of magnitude the measurement of $d_{e}$ using
a new technical method \cite{lamoreaux}. At BNL, a new experiment has been
proposed with the objective to reach a sensitivity of $10^{-24}e$ cm for $%
d_{\mu }$ \cite{semertzidis}. Neutrino factories or PRISM should be able to
reach a sensitivity of $10^{-26}e$ cm \cite{aysto}$.$ 

The lepton mixing matrix $U_{MNS}$ is also constrained by solar neutrino and
atmospheric neutrino data. Indeed, in case of LMA solution to the solar
neutrino problem, the mixing angles as defined in the standard
parametrisation for $U_{\delta }$ \cite{PDG} are typically in the ranges $%
0.24\lesssim \mbox{ tan}^{2}\theta _{12}\lesssim 0.89$~\cite{SNO}, $%
0.40\lesssim \mbox{ tan}^{2}\theta _{23}\lesssim 3.0$~\cite{SuperK}, and $%
\left| \mbox{ sin}\theta _{13}\right| \lesssim 0.2$~\cite{CHOOZ}. Here we
use the following values for the lepton mixing angles, $\theta _{12}=0.59$, $%
\theta _{23}=0.78$, and $\theta _{13}=0.2$ 

\subsection{Constraints from neutralino contributions}

First we consider the upper bounds on the relevant mass insertions in the
charged lepton sector, mediated by neutralino exchange. In Ref.\cite
{Gabbiani96} bounds on these mass insertions have been presented but only in
a very special case, where the lightest neutralino is assumed to be photino
like and of course, in that case, it also gives the dominant contribution to 
$C_{l_{i}}^{\chi ^{0}}$ and $A_{R,Lij}^{N}$. With these assumptions, the
bounds on the mass insertions depend only on the ratio $x=m_{\tilde{\gamma}%
}^{2}/m_{\tilde{l}}^{2}$, where $m_{\tilde{\gamma}}$ is the photino mass.
However, in a general SUSY model the bounds depend on the gaugino masses, $%
\mu $--term and $\tan \beta $. As we will show the bounds of some mass
insertions are sensitive to some of these parameters, in particular $\tan
\beta $. 

In table 2, we present the upper bounds on the absolute values of the mass
insertions $(\delta _{AB}^{l})_{12}$ (with $A,B=(L,R)$) from neutralino
contributions to the $BR(\mu \to e\gamma )$. We consider some representative
value of the ratio $x_{12}=M_{1}/M_{2}$ and fixed values of $\mu =\tilde{m}%
=200$ GeV, $M_{2}=100$ GeV and $\tan \beta =5$. 

\begin{center}
\begin{tabular}{|c||c|c|c|}
\hline
$x_{12}$ & $|(\delta _{LL}^{l})_{12}|$ & $|(\delta _{LR}^{l})_{12}|$ & $%
|(\delta _{RR}^{l})_{12}|$ \\ \hline
0.25 & $8.4\times 10^{-4}$ & $2.7\times 10^{-6}$ & $4.2\times 10^{-3}$ \\ 
\hline
0.5 & $1\times 10^{-3}$ & $1.8\times 10^{-6}$ & $1.7\times 10^{-3}$ \\ \hline
1 & $1.3\times 10^{-3}$ & $1.5\times 10^{-6}$ & $1.2\times 10^{-3}$ \\ \hline
2 & $1.4\times 10^{-3}$ & $1.8\times 10^{-6}$ & $1.8\times 10^{-3}$ \\ \hline
\end{tabular}
\end{center}

Table 2: Upper Bounds on $|(\delta _{LL}^{l})_{12}|$ from $BR(\mu \to
e\gamma )<1.2\times 10^{-11}$ for $\mu =\tilde{m}=200$ GeV, $M_{2}=100$ GeV,
and $\tan \beta =5$.\newline

From the result in table 2, it is remarkable that the strong
bounds on $|(\delta _{LR}^{l})_{12}|$ are the same for $ x_{12}$ and 
$1/x_{12}$, {\it i.e.}, they are insensitive to the nature of the lightest
neutralino, whether it is bino--like or wino--like. The dependence of the
absolute values of the mass insertions $(\delta _{AB}^{l})_{12}$ on $\tan
\beta $ is given in table 3 for $M_{1}=M_{2}=100$ GeV and $\mu =\tilde{m}=200
$ GeV. 

\begin{center}
\begin{tabular}{|c||c|c|c|}
\hline
$\tan \beta $ & $|(\delta _{LL}^{l})_{12}|$ & $|(\delta _{LR}^{l})_{12}|$ & $%
|(\delta _{RR}^{l})_{12}|$ \\ \hline
5 & $1.3\times 10^{-3}$ & $1.5\times 10^{-6}$ & $12\times 10^{-3}$ \\ \hline
15 & $6.5\times 10^{-4}$ & $1.5\times 10^{-6}$ & $6.7\times 10^{-4}$ \\ 
\hline
25 & $4.3\times 10^{-4}$ & $1.5\times 10^{-6}$ & $4.6\times 10^{-4}$ \\ 
\hline
35 & $3.3\times 10^{-4}$ & $1.5\times 10^{-6}$ & $3.6\times 10^{-4}$ \\ 
\hline
\end{tabular}
\end{center}

Table 3: Upper Bounds on $|(\delta _{LL}^{l})_{12}|$ from $BR(\mu \to
e\gamma )<1.2\times 10^{-11}$ for $\mu =\tilde{m}=200$ GeV and $%
M_{1}=M_{2}=100$ GeV.\newline

As can be seen from this table the bounds on the $LR$ mass insertion are
essentially independent on the values of $\tan \beta $. We have also found that the 
results are independent on the values $\mu $. Next, we consider the bounds on the 
imaginary parts of the relevant $LR$ mass insertions from the experimental limit 
of the electron EDM. As mentioned above, the $LL$ and $RR$ mass insertions are Hermitian, so
that $\mathrm{Im}(\delta _{LL,RR}^{l})_{ii}=0$ and only LR transitions
contribute to the EDM. In table 4, we present the upper bounds on $\mathrm{Im%
}(\delta _{LR}^{l})_{11}$ as function of the bino-wino ratio $x_{12}$ and $%
\mu $, from the experimental bound on electron EDM, $d_{e}<4.3\times 10^{-27}
$ e cm. 

\begin{center}
\begin{tabular}{|c||c|c|c|c|c|}
\hline
$x_{12}\;\;\mathbf{\backslash }\;\;\mu $ & 200 & 400 & 600 & 800 &  \\ 
\hline\hline
0.25 & $7.7\times 10^{-7}$ & $1.2\times 10^{-6}$ & $2.1\times 10^{-6}$ & $%
3.4\times 10^{-6}$ &  \\ \hline
0.5 & $8.4\times 10^{-7}$ & $1.1\times 10^{-6}$ & $2\times 10^{-6}$ & $%
3.3\times 10^{-6}$ &  \\ \hline
1 & $3.3\times 10^{-7}$ & $6.2\times 10^{-7}$ & $1.1\times 10^{-6}$ & $%
1.7\times 10^{-6}$ &  \\ \hline
2 & $6.8\times 10^{-7}$ & $1.4\times 10^{-7}$ & $2.4\times 10^{-6}$ & $%
4\times 10^{-6}$ &  \\ \hline
\end{tabular}
\end{center}

Table 4: Upper Bounds on $\mathrm{Im}(\delta _{LR}^{l})_{11}$ from electron
EDM, $d_{e}<4.3\times 10^{-27}$ e cm for $\tan \beta =5$, $\tilde{m}=200$
GeV and $M_{2}=100$ GeV.


\subsection{Constraints from chargino contributions}

Now we turn to the constraints on the LL mass insertion in the sneutrino
sector due to the chargino contributions to the LFV process $\mu \to e\gamma 
$ and the electron EDM. From eq.(\ref{chargino_mu_decays}) and using the
experimental bound given above for $Br(\mu \rightarrow e\gamma )$ together with 
the fact that $f_{4}(x)\ll f_{5}(x)$ for $x\simeq 1$, one can easily find a
limit on $\delta _{LL}^{\tilde{\nu}}$. Indeed, one gets 
\begin{eqnarray}
\left| \sum_{i,j}\left( \frac{100\mbox{ GeV}}{m_{a}\mbox{
cos}\beta }\right) U_{a2}^{*}V_{a1}^{*}(U_{MNS})_{ej}(\delta _{LL}^{\tilde{%
\nu}})_{ji}(U_{MNS}^{\dagger })_{i\mu }\right|  &=&6.4\times 10^{-4}\sqrt{%
\frac{Br(\mu \rightarrow e\gamma )}{1.2\times 10^{-11}}}  \nonumber \\
&\lesssim &6.4\times 10^{-4}  \label{bound on mu-e gamma}
\end{eqnarray}
As usually done in the mass insertion method, we assume that there is no
cancellation between contributions involving different $\delta _{LL}^{\tilde{%
\nu}}$ elements. So the bound given in eqs.(\ref{bound on mu-e gamma}) has
to be applied to each contribution $(\delta _{LL}^{\tilde{\nu}})_{ji}$. We
can proceed in the same way for $d_{e}$ and using the experimental limit on $%
d_{e}$, one gets 
\begin{eqnarray}
&&\mbox{ Im}\left( \sum_{i,j}\left( \frac{100\mbox{ GeV}}{m_{a}\mbox{
cos}\beta }\right) U_{a2}^{*}V_{a1}^{*}(U_{MNS})_{ej}(\delta _{LL}^{\tilde{%
\nu}})_{ji}(U_{MNS}^{\dagger })_{ie}\right)   \nonumber \\
&=&\sum_{i,j}\left( \frac{100\mbox{ GeV}}{m_{a}\mbox{
cos}\beta }\right) \mbox{ Im}\left( U_{a2}^{*}V_{a1}^{*}\right) \mbox{ Re}%
\left( (U_{MNS})_{ej}(\delta _{LL}^{\tilde{\nu}})_{ji}(U_{MNS}^{\dagger
})_{ie}\right)   \label{h1} \\
&=&2\times 10^{-2}\frac{d_{e}}{4.3\times 10^{-27}e\mbox{ cm}}  \nonumber \\
&\lesssim &2\times 10^{-2}  \label{edm bound}
\end{eqnarray}
where to get eqs(\ref{h1}), we use the hermiticity of $\delta _{LL}^{\tilde{%
\nu}}$. From eq.(\ref{edm bound}), it is clear that the chargino
contribution to the electron EDM do not lead to any significant constraint
on the LL mass insertions and as mentioned above the source of CP violation
in this case is the SUSY phase $\phi _{\mu }$. The most significant
constraints on the LL mass insertions come as usually from the $\mu \rightarrow
e\gamma $ experimental bound. In order to illustrate the dependence of the
 bounds given in eq. (\ref{bound on mu-e gamma}) on SUSY parameters, we
present in tables 5 and 6 the upper bounds on the magnitude of the relevant
LL mass insertions obtained from the experimental limits of LFV process $\mu \to
e\gamma $. To get these bounds, we used the values given in the beginning of
this section for the elements of $U_{MNS}$ lepton mixing matrix. 

\begin{center}
\begin{tabular}{|c||c|c|c|c|c|c|}
\hline
m & $|(\delta _{LL}^{\tilde{\nu}})_{11}|$ & $|(\delta _{LL}^{\tilde{\nu}%
})_{12}|$ & $|(\delta _{LL}^{\tilde{\nu}})_{13}|$ & $|(\delta _{LL}^{\tilde{%
\nu}})_{22}|$ & $|(\delta _{LL}^{\tilde{\nu}})_{23}|$ & $|(\delta _{LL}^{%
\tilde{\nu}})_{33}|$ \\ \hline
100 & $6.3\times 10^{-4}$ & $6.5\times 10^{-4}$ & $4.8\times 10^{-4}$ & $%
9.7\times 10^{-3}$ & $7.3\times 10^{-4}$ & $1.9\times 10^{-3}$ \\ \hline
200 & $6\times 10^{-4}$ & $6.2\times 10^{-4}$ & $4.6\times 10^{-4}$ & $%
8.1\times 10^{-4}$ & $6.8\times 10^{-4}$ & $1.8\times 10^{-3}$ \\ \hline
300 & $8\times 10^{-4}$ & $8.1\times 10^{-4}$ & $6.1\times 10^{-4}$ & $%
1.2\times 10^{-3}$ & $9\times 10^{-4}$ & $2.3\times 10^{-3}$ \\ \hline
400 & $1.1\times 10^{-3}$ & $1.2\times 10^{-3}$ & $8.6\times 10^{-4}$ & $%
1.4\times 10^{-3}$ & $1.3\times 10^{-4}$ & $3.4\times 10^{-3}$ \\ \hline
500 & $1.7\times 10^{-3}$ & $1.7\times 10^{-3}$ & $1.3\times 10^{-3}$ & $%
1.6\times 10^{-3}$ & $1.9\times 10^{-4}$ & $5\times 10^{-3}$ \\ \hline
\end{tabular}
\end{center}

\vspace{0.2cm} Table 5: Upper Bounds on $|(\delta _{LL}^{\tilde{\nu}})_{ij}|$
from $BR(\mu \to e\gamma )<1.2\times 10^{-11}$ for $M_{2}=\mu =200$ GeV and $%
\tan \beta =5$.\newline

\begin{center}
\begin{tabular}{|c||c|c|c|c|c|c|}
\hline
$\tan \beta $ & $|(\delta _{LL}^{\tilde{\nu}})_{11}|$ & $|(\delta _{LL}^{%
\tilde{\nu}})_{12}|$ & $|(\delta _{LL}^{\tilde{\nu}})_{13}|$ & $|(\delta
_{LL}^{\tilde{\nu}})_{22}|$ & $|(\delta _{LL}^{\tilde{\nu}})_{23}|$ & $%
|(\delta _{LL}^{\tilde{\nu}})_{33}|$ \\ \hline
5 & $6.3\times 10^{-4}$ & $6.7\times 10^{-4}$ & $5\times 10^{-4}$ & $%
9.9\times 10^{-3}$ & $7.3\times 10^{-4}$ & $1.7\times 10^{-3}$ \\ \hline
15 & $2.5\times 10^{-4}$ & $2.7\times 10^{-4}$ & $2\times 10^{-4}$ & $%
3.9\times 10^{-3}$ & $3\times 10^{-4}$ & $6.9\times 10^{-4}$ \\ \hline
25 & $1.6\times 10^{-4}$ & $1.7\times 10^{-4}$ & $1.2\times 10^{-4}$ & $%
2.5\times 10^{-3}$ & $1.8\times 10^{-4}$ & $4.4\times 10^{-4}$ \\ \hline
35 & $1.1\times 10^{-4}$ & $1.2\times 10^{-4}$ & $9\times 10^{-5}$ & $%
1.8\times 10^{-3}$ & $1.3\times 10^{-4}$ & $3\times 10^{-4}$ \\ \hline
45 & $9\times 10^{-5}$ & $1.4\times 10^{-5}$ & $7\times 10^{-5}$ & $%
9.9\times 10^{-3}$ & $1\times 10^{-4}$ & $2.5\times 10^{-4}$ \\ \hline
\end{tabular}
\end{center}

Table 6: Upper Bounds on $|(\delta _{LL}^{\tilde{\nu}})_{ij}|$ from $BR(\mu
\to e\gamma )<1.2\times 10^{-11}$ for $M_{2}=\mu =200$ GeV and $m=100$. 
\newline

Now, let us illustrate the relation between $Br(\mu \rightarrow e\gamma )$
and $d_{e}$ chargino contribution and what our experimental knowledge on
neutrino mixing matrix $U_{MNS}$ can tell us on relation between EDM's and
flavour changing processes. For that, one can define (without summation on
repeated indices) 
\begin{equation}
(U_{MNS})_{kj}(\delta _{LL}^{\tilde{\nu}})_{ji}(U_{MNS}^{\dagger
})_{il}\equiv \rho _{kl}^{ij}e^{i\varphi _{kl}^{ij}}
\end{equation}
In particular, one has 
\begin{equation}
\left| \rho _{e\mu }^{ij}\right| =\left| \rho _{ee}^{ij}\right| \left| \frac{%
(U_{MNS})_{i\mu }}{(U_{MNS})_{ie}}\right| 
\end{equation}
But as we can see from eqs(\ref{bound on mu-e gamma}), $\left| \rho _{e\mu
}^{ij}\right| $ 's are strongly constrained by $Br(\mu \rightarrow e\gamma )$%
. Using the experimental knowledge on $U_{MNS}$, one gets an upper limit on chargino
contribution to $d_{e}$, 
\begin{eqnarray}
\frac{d_{e}}{e} &\lesssim &1.37\times 10^{-28}\mbox{ sin}\phi _{\mu }\sqrt{%
\frac{Br(\mu \rightarrow e\gamma )}{1.2\times 10^{-11}}}\mbox{ cm} \\
\frac{d_{\mu }}{e} &\lesssim &2.82\times 10^{-26}\mbox{ sin}\phi _{\mu }%
\sqrt{\frac{Br(\mu \rightarrow e\gamma )}{1.2\times 10^{-11}}}\mbox{ cm}
\end{eqnarray}

Proceeding in the same way for $Br(\tau \rightarrow \mu \gamma )$, one has 
\begin{equation}
\left| \rho _{\mu \tau }^{ij}\right| =\left| \rho _{e\mu }^{ij}\right|
\left| \frac{(U_{MNS})_{i\tau }}{(U_{MNS})_{i\mu }}\right| \left| \frac{%
(U_{MNS})_{j\mu }}{(U_{MNS})_{je}}\right| 
\end{equation}
Due to atmospheric neutrino data, one has that 
\begin{equation}
\left| \rho _{\mu \tau }^{ij}\right| =\left| \rho _{e\mu }^{ij}\right|
\left| \frac{(U_{MNS})_{j\mu }}{(U_{MNS})_{je}}\right| 
\end{equation}
Thus the chargino contribution to $Br(\tau \rightarrow \mu \gamma )$ is
given by 
\begin{eqnarray}
Br(\tau \rightarrow \mu \gamma )\simeq \frac{1}{5}\frac{m_{\mu }^{2}}{%
m_{\tau }^{2}}Br(\mu \rightarrow e\gamma ) \\
\lesssim 7.8\times 10^{-15}  \nonumber
\end{eqnarray}

The same discussion can be done for $Br(\tau \rightarrow e\gamma )$  and one
gets similar results. 

\section{Neutrino $CP$ phases and EDM.}

In this section, we shall discuss the dependence of lepton EDM's on neutrino $CP$ 
phases appearing in $U_{MNS}$. It is clear that the running of the soft breaking terms 
from the GUT to the weak scale induce a dependence on $U_{MNS}$ (and particularly on its phases). 
However, their effects on the soft breaking terms are usually very small\footnote{ 
The effect of the running of the SUSY soft breaking terms from GUT to weak scale  
in the universal scenario for SUSY soft breaking terms has been recently studied in 
ref.\cite{ellis}.}. So, in the next discussion, we shall
neglect this dependence and we shall study two extreme cases. First, we
shall assume that all the SUSY soft-breaking terms are real and that the
only source of $CP$ violation arises from the charged lepton Yukawa
couplings. The second case corresponds to assuming that $\phi _{\mu }$, the
phase of the $\mu $-term, is different from zero and then checking 
what kind of textures for the SUSY soft-breaking terms may depend on
the neutrino Majorana phases.

Let us consider the case where there is no $CP$ violation coming
from the diagonalisation of the charginos and neutralinos mass matrices. In
this case, the lepton EDM has a very simple form, 
\begin{equation}
d_{l}/e=\frac{\alpha _{W}}{4\pi }\sum_{a=1}^{4}\frac{1}{m_{\chi _{a}^{0}}}%
f_{1}(x_{a})\mbox{ tan}\theta _{W}N_{a1}^{*}(N_{a2}^{*}+\mbox{ tan}\theta
_{W}N_{a1}^{*})~\mbox{ Im}(\delta _{RL}^{l})_{ii}
\end{equation}

At first sight, it seems to be independent of the low-energy neutrino $CP$
phases. But when $V_{L}^{\nu }\simeq \mathbf{1}$,   the effects of low
energy neutrino $CP$ phases  appear through the definition of $\delta
_{RL}^{l}$ . To illustrate this point, let us recall that for  $%
V_{L}^{\nu }\simeq \mathbf{1}$,  $U_{MNS}\simeq P_{L}V_{L}^{e\dagger }$. In
this case, the definition of $\delta _{RL}$ can be written as, 
\begin{eqnarray}
\delta _{RL} &\equiv &\frac{1}{\tilde{m}^{2}}v\cos \beta
~P_{L}(V_{R}^{e})^{\dagger }Y_{e}^{A}V_{L}^{e}P_{L}^{\dagger } \\
&=&\frac{1}{\tilde{m}^{2}}v\cos \beta ~U_{R}^{e}Y_{e}^{A}(U_{MNS})^{\dagger }
\end{eqnarray}
where to simplify the notation, we defined $U_{R}^{e}\equiv
P_{L}V_{R}^{e\dagger }$.  It is clear that the low energy neutrino $CP$
phases can strongly affect the EDM's through the definition of $Y_{e}^{A}$.
Indeed, one has 
\begin{eqnarray}
(Y_{e}^{A})_{ij} &\equiv &(A)_{ij}(Y_{e})_{ij}  \nonumber \\
&\equiv &(A)_{ij}(V_{R}^{e}\mbox{ diag}(h_{e},h_{\mu },h_{\tau
})V_{L}^{e\dagger })_{ij} \\
&=&(A)_{ij}(U_{R}^{e\dagger }\mbox{ diag}(h_{e},h_{\mu },h_{\tau
})U_{MNS})_{ij}
\end{eqnarray}
To get the last line, we used the fact that $P_{L}$ is a diagonal unitary
matrix and commutes with $\mbox{ diag}(h_{e},h_{\mu },h_{\tau })$. This
result depends on the texture for the trilinear terms $(A)_{ij}$, but it is possible to extract
some general properties. Indeed, any $A$ matrix can be written as follows, 
\begin{equation}
A\equiv a\left( 
\begin{array}{ccc}
1\; & 1 & 1 \\ 
1\; & 1 & 1 \\ 
1\; & 1 & 1
\end{array}
\right) \ +\sum_{p,q}c^{pq}b_{pq}
\end{equation}
where $b_{pq}$ is define as a matrix with all entries equal to zero except
for the element $(p,q)$ which is equal to $1$ and $c^{pq}$ are numerical
coefficients. The first term corresponds to the usual universal trilinear
terms where $Y_{e}^{A}=aY^{e}$.  Im$(\delta _{RL}^{l})_{11}$ can now be
rewritten as 
\begin{eqnarray}
\mbox{ Im}(\delta _{RL}^{l})_{11} &=&\sum_{p,q,j=1}^{3}\frac{1}{\tilde{m}^{2}%
}v\cos \beta \mbox{ Im}\left(
c^{pq}(U_{R}^{e})_{1p}(U_{R}^{e*})_{jp}h_{jj}(U_{MNS})_{jq}(U_{MNS}^{*})_{1q}\right) 
\\
&=&\sum_{p,q,j=1}^{3}\frac{\sqrt{2}m_{j}}{\tilde{m}^{2}}\mbox{ Im}~\left(
c^{pq}(U_{R}^{e})_{1p}(U_{R}^{e*})_{jp}(U_{MNS})_{jq}(U_{MNS}^{*})_{1q}%
\right)   \label{Imdelta_b}
\end{eqnarray}
with $m_{j}$, the charged lepton masses ($m_{1,2,3}=m_{e,\mu ,\tau }$). We
can directly see from eqs(\ref{Imdelta_b}) that with universal trilinear couplings,
the $\mbox{ Im}(\delta_{RL}^{l})_{11}$ is independent of the Majorana or Dirac neutrino phases. But
for a  texture for the $A$ matrix different from the universal case, $\mbox{
Im}(\delta _{RL}^{l})_{11}$ depends on the neutrino Dirac phase.  

An interesting limit is to consider the case of Hermitian Yukawa coupling
for the charged leptons ($U_{MNS}=U_{R}^{e}$). In that case, eq.(\ref
{Imdelta_b}) reads as 
\begin{equation}
\mbox{ Im}(\delta _{RL}^{l})_{11}=\sum_{p,q,j=1}^{3}\frac{\sqrt{2}m_{j}}{%
\tilde{m}^{2}}\mbox{ Im}\left(
c^{pq}~(U_{MNS})_{1p}(U_{MNS}^{*})_{jp}(U_{MNS})_{jq}(U_{MNS}^{*})_{1q}%
\right) 
\end{equation}
It is important to notice that, in the Hermitian case,  there is no
contribution to Im$(\delta _{RL})$ coming from  the Dirac $CP$ violating
phase of the mixing matrix $U_{MNS}$   for $p=q$, and that the $A$ matrix
has to be not Hermitian. Otherwise, $\mbox{ Im}(\delta _{RL}^{l})_{11}=0$.

In case of no $CP$ violation coming from the $A$ matrix (all $c^{pq}$ are
real~\footnote{Note that even if the trilinear couplings are real at GUT scale, they 
receive small complex contributions from the complex Yukawa through the running to the electrweak
scle. Here we neglect this effect.}), one has
\begin{eqnarray}
\mbox{ Im}(\delta _{RL}^{l})_{11} &=&\sum_{p,q,j=1}^{3}\frac{\sqrt{2}m_{j}}{%
\tilde{m}^{2}}c^{pq}\mbox{ Im}\left(
(U_{MNS})_{1p}(U_{MNS}^{*})_{jp}(U_{MNS})_{jq}(U_{MNS}^{*})_{1q}\right) 
\label{rephasing_invariant} \\
&\simeq &\sum_{p,q=1}^{3}\frac{\sqrt{2}m_{\tau }}{\tilde{m}^{2}}c^{pq}\mbox{
Im}\left(
(U_{MNS})_{1p}(U_{MNS}^{*})_{3p}(U_{MNS})_{3q}(U_{MNS}^{*})_{1q}\right) 
\end{eqnarray}
where the imaginary part which appears in eq.(\ref{rephasing_invariant}) is
the usual rephasing invariant measure of Dirac $CP$
violation\footnote{%
By Dirac $CP$ violation, we mean $CP$ violation arising from the
Dirac phase of the $U_{MNS}$ lepton mixing matrix.}. Indeed, in neutrino
oscillations, the $CP$ asymmetry defined as the difference of the $CP$
conjugated neutrino oscillation probabilities $P(\nu _{e}\rightarrow \nu
_{\mu })-P(\overline{\nu }_{e}\rightarrow \overline{\nu }_{\mu })$ is
proportionnal to the imaginary part of an invariant quartet $\mathcal{J}_{CP}
$ defined as
\begin{equation}
\mathcal{J}_{CP}\equiv \mbox{ Im}\left(
(U_{MNS})_{11}(U_{MNS}^{*})_{21}(U_{MNS})_{22}(U_{MNS}^{*})_{12}\right) 
\end{equation}
As all the rephasing invariant quartets are equal up to their sign, $\mbox{
Im}(\delta _{RL}^{l})_{11}$ can be written as
\begin{eqnarray}
\left| \mbox{ Im}(\delta _{RL}^{l})_{11}\right|  &\simeq &\left| \mathcal{J}%
_{CP}\right| \left| \sum_{p,q=1;p\neq q}^{3}\frac{\sqrt{2}%
m_{\tau }}{\tilde{m}^{2}}c^{pq}\right|   \label{JCP1} \\
&\lesssim &10^{-6}  \label{JCP2}
\end{eqnarray}
where the last inequality is obtained using table 4. This means that if the
lepton large mixings and $CP$ violation  has its origin in the charged
leptonYukawa coupling, in case of Hermitian Yukawa coupling and for a given 
texture of the trilinear $A$ terms, one has a simple correlation between the
measure of electron EDM and $CP$ asymmetries in neutrino oscillation. For
instance, assuming that $c^{pq}\sim \tilde{m}\sim v$, using eqs.(\ref{JCP1}-%
\ref{JCP2}), one gets a limit on  $\left| \mathcal{J}_{CP}\right| ,$%
\begin{equation}
\left| \mathcal{J}_{CP}\right| \lesssim 7\times 10^{-5}  \label{JCP3}
\end{equation}
It is amazing to note that this value is very close to the experimental
measure of the rephasing invariant quartet of the quark sector, $\left| 
\mathcal{J}_{CP}^{q}\right| $. Indeed, one has\cite{PDG},
\begin{equation}
\left| \mathcal{J}_{CP}^{q}\right| =\left( 3.0\pm 0.3\right) \times 10^{-5}.
\end{equation}


Before concluding, let us discuss the case where the $CP$ violation
arise from both the Yukawa couplings and the SUSY parameters, in particlar, 
if the diagonalisation of the chargino or the neutralino mass matrices. As an illustrative example, we
shall discuss the case of the chargino contribution. The discussion for the
neutralino contribution can be extended in a straigthforward way. In this case, the
chargino contribution to electron EDM is given by 
\begin{equation}
\frac{d_{e}}{4.3\times 10^{-27}e\mbox{ cm}}=50\times \sum_{i,j}\left( \frac{%
100\mbox{ GeV}}{m_{a}\mbox{
cos}\beta }\right) \mbox{ Im}\left( U_{a2}^{*}V_{a1}^{*}\right) \mbox{ Re}%
\left( (U_{MNS})_{ej}(\delta _{LL}^{\tilde{\nu}})_{ji}(U_{MNS}^{\dagger
})_{ie}\right) 
\end{equation}
The neutrino $CP$ violating phases dependence can appear directly through $%
U_{MNS}$ but also through the definition of $\delta _{LL}^{\tilde{\nu}}$ as
given by eqs.(\ref{sqmass2}). In case of $V_{L}^{\nu }\simeq \mathbf{1}$,
one has $(\delta _{LL}^{\tilde{\nu}})_{ji}\simeq (\widetilde{m}%
_{l}^{2})_{ji}/\widetilde{m}^{2}$. By neglecting the Dirac CP-violating phase of $U_{MNS}$ and using
eqs(11), one gets a simple expression for the EDM in terms of the Majorana
CP-violating phases,
\begin{eqnarray}
\frac{d_{e}}{4.3\times 10^{-27}e\mbox{ cm}} &=&50\times \sum_{i,j}\left( 
\frac{100\mbox{ GeV}}{m_{a}\mbox{
cos}\beta }\right) \mbox{ Im}\left( U_{a2}^{*}V_{a1}^{*}\right) \times  
\nonumber \\
&&\mbox{ cos}\left( \phi _{j}^{M}-\phi _{i}^{M}\right) \mbox{ Re}\left(
(U_{\delta })_{ej}(\delta _{LL}^{\tilde{\nu}})_{ji}(U_{\delta }^{\dagger
})_{ie}\right)   \label{de_majorana}
\end{eqnarray}
It is clear from eq.(\ref{de_majorana}) that, in the case of a texture for  $(\delta
_{LL}^{\tilde{\nu}})_{ji}$ different from the universal case, the electron
EDM is a function of the cosinus of the Majorana phases and could be used as
a way to probe the Majorana phases for a given texture for the SUSY
soft-breaking terms. 

\section{Conclusion}

In this paper, we have analyzed the constraints obtained from the chargino 
and neutralino contributions to the lepton EDM and LFV. We have adopted the 
mass insertion method which implies a model independent parametrization.
We have provided analytical results for these contributions as functions
of the leptonic mass insertions. We also derived model independent upper
bounds on the relevant mass insertions by requiring that the pure chargino 
or neutralinio contribution do not exceed the experimental limit of the 
lepton EDM and LFV (in particular the electron EDM and the branching ratio
of $\mu \to e \gamma$, which give the most strangent bounds).

It was emphasized that once the bounds from $\mu \to e \gamma$ are impossed, 
the current experimental limits on the EDMs can be satisfied for any value
of neutrino phases, whatever is the structure of the soft SUSY breaking 
terms.

\section*{Acknowledgements}
This work was partially supported by the Nato Collaborative Linkage Grants. The work of D.D.
was supported by \emph{Funda\c{c}\~{a}o para a Ci\^{e}ncia e a Tecnologia}
(FCT) through the project POCTI/36288/FIS/2000. The work of S.K. was supported by
PPARC.



\begin{thebibliography}{99}

\bibitem{SuperK} 
S.~Fukuda {\it et al.}  [Super-Kamiokande Collaboration],
Phys.\ Rev.\ Lett.\  {\bf 86} (2001) 5656
[arXiv:hep-ex/0103033].
S.~Fukuda {\it et al.}  [Super-Kamiokande Collaboration],
Phys.\ Rev.\ Lett.\  {\bf 86} (2001) 5651
[arXiv:hep-ex/0103032].
S.~Fukuda {\it et al.}  [Super-Kamiokande Collaboration],
Phys.\ Lett.\ B {\bf 539} (2002) 179
[arXiv:hep-ex/0205075].


\bibitem{CHOOZ} 
M.~Apollonio {\it et al.}  [CHOOZ Collaboration],
Phys.\ Lett.\ B {\bf 466} (1999) 415
[arXiv:hep-ex/9907037].


\bibitem{SNO} 
Q.~R.~Ahmad {\it et al.}  [SNO Collaboration],
Phys.\ Rev.\ Lett.\  {\bf 87} (2001) 071301
[arXiv:nucl-ex/0106015].
Q.~R.~Ahmad {\it et al.}  [SNO Collaboration],
Phys.\ Rev.\ Lett.\  {\bf 89} (2002) 011301
[arXiv:nucl-ex/0204008].

\bibitem{MACRO}  M.~Ambrosio \textit{et al.} [MACRO Collaboration], 
Phys.\ Lett.\ B \textbf{517} (2001) 59; 
G.~Giacomelli and M.~Giorgini [MACRO Collaboration], 
arXiv:hep-ex/0110021. 

\bibitem{KAMLAND}
K. Eguchi et al. [KamLAND Collaboration],
arXiv:hep-ex/0212021.




\bibitem {SM}G.~R.~Farrar and M.~E.~Shaposhnikov,
Phys.\ Rev.\ Lett.\ \textbf{70}, 2833 (1993) [Erratum, ibid.\ \textbf{71},
210 (1993)]; M.~B.~Gavela, P.~Hernandez, J.~Orloff, O.~Pene and
C.~Quimbay, Nucl.\ Phys.\ B \textbf{430}, 382 (1994); M.~B.~Gavela,
M.~Lozano, J.~Orloff and O.~Pene, Nucl.\ Phys.\ B \textbf{430}, 345
(1994); P.~Huet and E.~Sather, Phys.\ Rev.\ D \textbf{51}, 379 (1995).


\bibitem{lightstop}
D.~Delepine, J.~M.~Gerard, R.~Gonzalez Felipe and J.~Weyers,
Phys.\ Lett.\ B {\bf 386} (1996) 183
[arXiv:hep-ph/9604440];
D.~Delepine,
arXiv:hep-ph/9609346;
M.~Carena, M.~Quiros and C.~E.~Wagner,
Phys.\ Lett.\ B \textbf{380}, 81 (1996);  M.~Carena, M.~Quiros and
C.~E.~Wagner, Nucl.\ Phys.\ B \textbf{524}, 3 (1998);
J.~R.~Espinosa, Nucl.\ Phys.\ B
\textbf{475}, 273 (1996);
J.~M.~Cline and K.~Kainulainen,
Nucl.\ Phys.\ B \textbf{482}, 73 (1996); J.~M.~Cline, M.~Joyce and
K.~Kainulainen, Phys.\ Lett.\ B \textbf{417}, 79 (1998)  [Erratum, ibid.\
B \textbf{448}, 321 (1999)].


\bibitem{edm1}
S.~Pokorski, J.~Rosiek and C.~A.~Savoy, Nucl.\ Phys.\ B {\bf 570}, 81
(2000);
 S.~Abel, S.~Khalil and O.~Lebedev,
Nucl.\ Phys.\ B \textbf{606}, 151 (2001) and references therein;
D.~Chang, W.~Y.~Keung and A.~Pilaftsis,
Phys.\ Rev.\ Lett.\ \textbf{82}, 900 (1999) [Erratum, ibid.\ \textbf{83},
3972 (1999)]; D.~Chang, W.~F.~Chang and W.~Y.~Keung, Phys.\ Lett.\ B
\textbf{478}, 239 (2000); D.~Chang, W.~F.~Chang and W.~Y.~Keung,
\texttt{hep-ph/0205084};
 A. Pilaftsis, \texttt{hep-ph/0207277}.


\bibitem{ana}
D.~Delepine, R.~Gonzalez Felipe, S.~Khalil and A.~M.~Teixeira,
Phys.\ Rev.\ D {\bf 66} (2002) 115011
[arXiv:hep-ph/0208236].

\bibitem{leptogenesis}
M.~Fukugita and T.~Yanagida,
Phys.\ Lett.\ B {\bf 174} (1986) 45;
L.~Covi, E.~Roulet and F.~Vissani,
Phys.\ Lett.\ B {\bf 384} (1996) 169
[arXiv:hep-ph/9605319].
M.~Flanz, E.~A.~Paschos and U.~Sarkar,
Phys.\ Lett.\ B {\bf 345} (1995) 248
[Erratum-ibid.\ B {\bf 382} (1996) 447]
[arXiv:hep-ph/9411366];
W.~Buchmuller and M.~Plumacher,
Phys.\ Lett.\ B {\bf 389} (1996) 73
[arXiv:hep-ph/9608308].

\bibitem{cplowenergy-lepto}
G.~C.~Branco, L.~Lavoura and M.~N.~Rebelo,
Phys.\ Lett.\ B {\bf 180} (1986) 264;
A.~Pilaftsis,
Phys.\ Rev.\ D {\bf 56} (1997) 5431
[arXiv:hep-ph/9707235];
G.~C.~Branco, T.~Morozumi, B.~M.~Nobre and M.~N.~Rebelo,
Nucl.\ Phys.\ B {\bf 617} (2001) 475
[arXiv:hep-ph/0107164];
W.~Rodejohann and K.~R.~Balaji,
Phys.\ Rev.\ D {\bf 65} (2002) 093009
[arXiv:hep-ph/0201052];
G.~C.~Branco, R.~Gonzalez Felipe, F.~R.~Joaquim and M.~N.~Rebelo,
Nucl.\ Phys.\ B {\bf 640} (2002) 202
[arXiv:hep-ph/0202030];
J.~R.~Ellis and M.~Raidal,
Nucl.\ Phys.\ B {\bf 643} (2002) 229
[arXiv:hep-ph/0206174];
S.~Davidson and A.~Ibarra,
arXiv:hep-ph/0206304;
P.~H.~Frampton, S.~L.~Glashow and T.~Yanagida,
Phys.\ Lett.\ B {\bf 548} (2002) 119
[arXiv:hep-ph/0208157];
T.~Endoh, S.~Kaneko, S.~K.~Kang, T.~Morozumi and M.~Tanimoto,
Phys.\ Rev.\ Lett.\  {\bf 89} (2002) 231601
[arXiv:hep-ph/0209020];
G.~C.~Branco {\it et al.},
arXiv:hep-ph/0211001;

\bibitem{lamoreaux}
S.~K.~Lamoreaux,
arXiv:nucl-ex/0109014.

\bibitem{Gabbiani96}  F.~Gabbiani, E.~Gabrielli, A.~Masiero and
L.~Silvestrini, 
Nucl.\ Phys.\ B \textbf{477}, 321 (1996) [arXiv:hep-ph/9604387]. 

\bibitem{Khalil:2001wr}
S.~Khalil, T.~Kobayashi and A.~Masiero,
Phys.\ Rev.\ D {\bf 60}, 075003 (1999);
S.~Khalil and O.~Lebedev,
Phys.\ Lett.\ B {\bf 515}, 387 (2001);
E.~Gabrielli and S.~Khalil,
Phys.\ Rev.\ D {\bf 67}, 015008 (2003).

\bibitem{ellis}
J.~R.~Ellis, J.~Hisano, M.~Raidal and Y.~Shimizu,
Phys.\ Lett.\ B {\bf 528} (2002) 86
[arXiv:hep-ph/0111324].



\bibitem{seesaw}  M. Gell-Mann, P. Ramond and R. Slansky, in Supersymmetry,
eds. D. Freedman and P. van Nieuwenhuizen North Holland, Amsterdam, p.315
(1979); T. Yanagida, in Proceedings of the workshop on unified theory and
baryon number in the universe, eds. O. Sawada and A.\ Sugamoto, KEK,
Tsukuba, Japan (1979); R.~N.~Mohapatra and G.~Senjanovic, 
Phys.\ Rev.\ Lett.\ 44 (1980) 912; 
R.~N.~Mohapatra and G.~Senjanovic, 
Phys.\ Rev.\ D 23 (1981) 165. 

\bibitem{vissani98}
see {\it e.g.},
P.~Binetruy, S.~Lavignac and P.~Ramond,
Nucl.\ Phys.\ B {\bf 477} (1996) 353
[arXiv:hep-ph/9601243];
M.~E.~Gomez, G.~K.~Leontaris, S.~Lola and J.~D.~Vergados,
Phys.\ Rev.\ D {\bf 59} (1999) 116009
[arXiv:hep-ph/9810291];
S.~Lola and G.~G.~Ross,
Nucl.\ Phys.\ B {\bf 553} (1999) 81
[arXiv:hep-ph/9902283];
W.~Buchmuller, D.~Delepine and F.~Vissani,
Phys.\ Lett.\ B {\bf 459} (1999) 171
[arXiv:hep-ph/9904219];
W.~Buchmuller, D.~Delepine and L.~T.~Handoko,
Nucl.\ Phys.\ B {\bf 576} (2000) 445
[arXiv:hep-ph/9912317].
 



\bibitem{Hisano96}
J.~Hisano and D.~Nomura,
Phys.\ Rev.\ D {\bf 59} (1999) 116005
[arXiv:hep-ph/9810479].

\bibitem{commins1994}
E.~D.~Commins, S.~B.~Ross, D.~DeMille and B.~C.~Regan,
Phys.\ Rev.\ A {\bf 50} (1994) 2960.


\bibitem{g-2}
H.~N.~Brown {\it et al.}  [Muon g-2 Collaboration],
Phys.\ Rev.\ Lett.\  {\bf 86} (2001) 2227
[arXiv:hep-ex/0102017].
G.~W.~Bennett {\it et al.}  [Muon g-2 Collaboration],
Phys.\ Rev.\ Lett.\  {\bf 89} (2002) 101804
[Erratum-ibid.\  {\bf 89} (2002) 129903]
[arXiv:hep-ex/0208001].


 

\bibitem{acciarri1998}
M.~Acciarri {\it et al.}  [L3 Collaboration],
Phys.\ Lett.\ B {\bf 434} (1998) 169.
 

\bibitem{mega}
M.~L.~Brooks {\it et al.}  [MEGA Collaboration],
Phys.\ Rev.\ Lett.\  {\bf 83} (1999) 1521
[arXiv:hep-ex/9905013].


 

\bibitem{CLEO}
S.~Ahmed {\it et al.}  [CLEO Collaboration],
Phys.\ Rev.\ D {\bf 61} (2000) 071101
[arXiv:hep-ex/9910060].


 
\bibitem{PDG}
K.~Hagiwara {\it et al.}  [Particle Data Group Collaboration],
Phys.\ Rev.\ D {\bf 66} (2002) 010001.
Rev.\ D 66 (2002) 010001. 

\bibitem{PSI}  L.M. Barkov et al., Research Proposal for experiment at PSI,\
1999. 

\bibitem{aysto}
J.~Aysto {\it et al.},
arXiv:hep-ph/0109217.
 



\bibitem{semertzidis}
Y.~K.~Semertzidis {\it et al.},
arXiv:hep-ph/0012087.
 

\bibitem{savoy}I. Masina, C.A. Savoy, hep-ph/0211283










\end{thebibliography}
\end{document}